# Current-induced spin polarization in topological insulator-graphene heterostructures

*Kristina Vaklinova[1*], Alexander Hoyer[1], Marko Burghard[1] and Klaus Kern[1,2]*

[1]Max Planck Institute for Solid State Research, Heisenbergstrasse 1, D-70569 Stuttgart, Germany

[2]Institut de Physique de la Matière Condensée, Ecole Polytechnique Fédérale de Lausanne, CH-1015 Lausanne, Switzerland

**ABSTRACT:**

Further development of the field of all-electric spintronics requires the successful integration of spin transport channels with spin injector/generator elements. While with the advent of graphene and related 2D materials high performance spin channel materials are available, the use of nanostructured spin generators remains a major challenge. Especially promising for the latter purpose are 3D topological insulators, whose 2D surface states host massless Dirac fermions with spin-momentum locking. Here, we demonstrate injection of spin-polarized current from a topological insulator into graphene, enabled by its intimate coupling to an ultrathin $Bi_2Te_2Se$ nanoplatelet within a van der Waals epitaxial heterostructure. The spin switching signal, whose magnitude scales inversely with temperature, is detectable up to ~15 K. Our findings establish topological insulators as prospective future components of spintronic devices wherein spin manipulation is achieved by purely electrical means.

The major goal of spintronics is to achieve efficient control over spin current and spin configurations by electrical and/or optical means[1,2]. Spintronic devices comprise two essential components, namely spin-transport and spin-injector/generator elements. The first component is usually implemented by conductors with weak spin-orbit (SO) coupling, enabling them to transmit spin-encoded information across the device. One prominent example is graphene, an effective spin conserver featuring spin diffusion lengths of up to several μm at room temperature[3,4]. Although the detailed mechanism that limits spin lifetimes in graphene is still debated, recent experimental and theoretical studies point toward the importance of spin-flip scattering by magnetic moments (originating from vacancies or adatoms)[5,6]. Spin injectors as the second component are typically ferromagnetic metal contacts (e.g., Co or permalloy $Ni_{0.8}Fe_{0.2}$). Achieving efficient spin injection is a challenging task, requiring a tailored interface between the spin channel material and the ferromagnetic metal. Notable progress in this direction has recently been made with the aid of homoepitaxial tunnel barrier contacts[7] and hBN tunnel barriers[8]. However, due to complications like stray field effects, arising in the downscaling process or due to roughness of the ferromagnetic interface[9], spintronic devices which operate without ferromagnetic contacts have attracted increasing attention[10].

One strategy to replace ferromagnetic contacts involves spin filter tunneling through ultrathin magnetic oxides. For instance, epitaxial monolayers of magnetic europium oxide provide a promising route to spin contacts directly on silicon[11,12]. An alternative approach comprises paramagnetic conductors with strong SO coupling, which imparts spin-generator capability. Promising candidates are three-dimensional (3D) topological insulators (TIs), which behave as insulators in the bulk but have topologically protected surface states described by the two-dimensional (2D) Dirac equation[13]. In general, 3D TIs lack intrinsic magnetic order, although theory indicates the possibility of collective alignment of dilute moments deliberately introduced into the system[14]. The ability of 3D TIs to generate pure spin-polarized currents arises from the spin-momentum locking of the 2D surface states, with the amplitude and direction of the charge current governing the amplitude and direction of the resulting net spin polarization. Charge current-induced spin-polarization in the 3D TIs $Bi_2Se_3$ or $Bi_2Te_2Se$ has been electrically detected with the aid of ferromagnetic contacts in different configurations[15-21]. Moreover, it has been experimentally shown that charge current flowing within a 3D TI can exert spin-transfer torque onto an adjacent ferromagnetic metal film[22]. By contrast, the injection of a net spin current from a TI into a nearby spin channel material, while theoretically predicted[23], has not yet been experimentally realized. According to theoretical studies, combining a 3D TI with a suitable 2D material such as graphene could enable transferring topological insulator properties across the interface, which might prove useful for engineering novel types of spintronic devices[24-26].

Here, we experimentally demonstrate graphene-based spin valves in which spin injection occurs from an ultrathin layer of the 3D TI bismuth selenide telluride $Bi_2Te_2Se$ (BTS), deposited on top of the graphene sheet by van der Waals epitaxy. Epitaxial growth is facilitated by the small lattice constant mismatch of about 1.5%[27] and ensures a close coupling between the two materials, as concluded from an increased carrier mobility of BTS nanoplatelets on hBN sheets[28], and corresponding transmission electron microscopy analysis[29]. The compound BTS is advantageous due to its relatively high bulk resistivity[30], large band gap of 310 meV, and reduced n-doping in comparison to $Bi_2Se_3$.

After mechanical exfoliation of highly oriented pyrolytic graphite (HOPG) onto Si substrates covered by a 300 nm thick $SiO_2$ layer, $Bi_2Te_2Se$ (BTS) nanoplatelets were grown on top by a catalyst-free vapor-solid process. To this end, $Bi_2Se_3$ and $Bi_2Te_3$ crystal sources (Alfa Aesar, 99.999%) were placed in a tube furnace, close to the homogeneous hot zone. The graphene-covered Si/$SiO_2$ growth substrates were placed in the colder region, where the temperature gradient can be exploited to adjust the substrate temperature as an important growth parameter. The quartz tube was then evacuated to 80 mbar and the temperature ramped up to 582°C using well-defined heating rates. As the sources begin to evaporate between 450 and 470°C a continuous ultrapure Argon flow (6N) of 150 sccm carries the evaporated material over to the deposition substrates where the crystals are formed.

In a first lithography step, Ti/Au markers were fabricated and the graphene-TI stacks studied with AFM to determine height profiles and surface morphology. All further patterning was done via e-beam lithography using PMMA resist. The entire device fabrication included 4 major steps, specifically: i) reactive ion etching to pattern the graphene into strips with widths of 200-300 nm in order to increase the resistance of the graphene/Co interface; ii) thermal evaporation of a 25 nm-thick $SiO_x$ film on a small area covering the edge of the graphene/TI stack, such that later only the TI can be contacted; iii) pretreatment of the TI contact regions inside a deposition chamber (base pressure of $3 \cdot 10^{-8}$ mbar) by 15 sec of argon milling at $10^{-4}$ mbar in order to remove the surface oxide layer and achieve Ohmic contacts. Subsequently, 2 nm Ti/40 nm Au non-magnetic contacts were thermally evaporated at $8 \cdot 10^{-8}$ mbar; iv) definition of ferromagnetic contacts on the graphene strips by thermal evaporation of 45 nm Co, followed by 10 nm Au as a capping layer to prevent Co oxidation. Prior to the Co/Au deposition, the chamber was evacuated for 2 days and the sample heated in order to minimize (surface) contaminations and reach the base pressure of the chamber.

Pure metallic Co structures were fabricated in order to confirm the material's ferromagnetic property, as probed by low temperature magnetotransport measurements. The graphene/Co contact resistance was found to fall into the range of 1-5 k$\Omega$, whereas the non-magnetic TI/Au contacts displayed values between 200 and 700 $\Omega$. The contact resistances in both case changed only little upon cooling from RT to 1.3 K.

Transport measurements were performed in an Oxford cryostat equipped with a 12 T magnet and a rotatable variable range insert for measurements in the range of 1.3 - 300 K. The applied 13 Hz AC-signal is offset by a home-built differential operational amplifier module running on batteries. The current is monitored by a DLPCA 200 (FEMTO) and the non-local voltage is pre-amplified using a SRS 560 (Stanford) before digitalization. In total, five devices were characterized for three of which a complete data set was obtained.

The spin valve measurements were carried out by passing a constant AC (1 μA) mixed with a DC (±5 to ±20 μA) current through the non-magnetic contacts on top of the BTS and measuring the voltage drop between a pair of ferromagnetic contacts as a function of in-plane magnetic field applied along their easy axis of magnetization perpendicular to the applied bias current.

The device configuration used to detect the spins injected into graphene is schematically illustrated in Fig. 1a. Importantly, the placement of one (non-magnetic) electrode pair on the BTS combined with at least two (magnetic) electrodes on the graphene allows for direct comparison between spin injection from a ferromagnetic cobalt contact and the TI. Moreover, this configuration allows for non-local 4-terminal measurements, which provide more reliable information compared to local measurements of spin polarization in 3D TIs[31,32]. Part of the bare graphene is patterned into a narrow strip with width between 200 and 300 nm, in order to reduce the contact area between the Co electrode and graphene. The resulting increase of contact resistance renders the strongly coupled (transparent) Co contacts suitable as spin probes in the absence of a tunneling barrier[33]. The electrical transport behavior presented below was reproducibly observed for five different devices.

By using catalyst-free vapor-solid growth, we obtained regularly shaped BTS nanoplatelets on mechanically exfoliated graphene, as exemplified in Fig. 1b. The lateral sizes of as-grown platelets range from 400 nm to 5 μm, depending on the shape of the graphene underneath, while their thickness is between 4 and 15 nm. The growth time is adjusted such that the platelets are still well-separated from each other. In separate magnetotransport experiments on individual BTS nanoplatelets grown under similar conditions directly on Si/SiO$_2$ substrates, we observed weak antilocalization (WAL) features attributable to the 2D surface state of BTS for platelet thicknesses above 6 nm.

Our first set of experiments addresses spin transport in the narrowed section of the underlying graphene sheet (Fig. 2). In the used geometry, spin polarized carriers are injected by a Co electrode and detected by a nearby Co contact in a 4-terminal, non-local measurement (Fig. 2a). The magnetization of both electrodes is governed by the externally applied in-plane B-field. The devices display a classical spin valve effect in the graphene channel, which manifests itself as a change in the non-local voltage/resistance, when the applied current is kept constant (Fig. 2b). Specifically, the resistance is low when the magnetic injector and detector are magnetized parallel to each other. As one of them reverses its magnetization direction (at a coercive field of 50 mT for an electrode width of 170 nm), the resulting antiparallel alignment causes a sharp resistance increase. Due to the similar lateral width of the injector and detector Co electrodes, the high-resistance plateau is quite narrow, which leads to a peak-like appearance. The sign of the non-local resistance, and therefore the orientation of the switching steps, depends on the sign of the applied bias. No background has been subtracted from the raw data. As a consequence of the suppressed electron-phonon scattering in graphene, the spin signal changes only little up to 40 K (Fig. 2c). This behavior is in agreement with previous reports on graphene-based spin valves up to room temperature[34]. Back gate-dependent measurements (Fig. 2d) reveal a minimum of the non-local signal $\Delta R_{NL}$ at the charge neutrality point (CNP) which occurs at $V_g = +32$ V for the device shown in Fig. 2. This behavior signifies the presence of transparent ferromagnetic contacts[35], as expected from the direct evaporation of Co onto the graphene strip and can be explained within the one-dimensional (1D) drift-diffusion theory of spin transport[36]. It is furthermore noteworthy that the relatively low resistance at the CNP (8 kΩ) reflects (inhomogeneous) doping of the graphene by PMMA residues accumulated during the device fabrication process[37] and allows for spin detection even at the CNP. For the field-effect mobility, values in the range of 1250-2450 cm$^2$V$^{-1}$s$^{-1}$ are extracted from the graphene transfer curves.

Having established the conventional spin valve operation in the bare graphene section, in the second set of experiments we inject current between two non-magnetic Ti/Au electrodes on top of the graphene/BTS heterostructure and detect the spin signal between a pair of ferromagnetic electrodes on the graphene strip (Fig. 3a). Again B-field induced resistance switching is observed in the form of a hysteretic step-like voltage signal (Fig. 3c,d). However, in comparison to Fig. 2b, only a single-switch is observed. This difference can be explained by the fact that due to the spin-momentum locking of the BTS surface states, the two spin directions parallel to the magnetization axis of the ferromagnetic electrodes are locked to the two momentum vectors, $k_x$ and $-k_x$, corresponding to positive and negative bias current. For the present devices, this mechanism is confirmed by the effect of reversing the polarity of current injected into the BTS. For positive bias current, $I_{dc}$ = +5 µA, the resistance change is consistent with the detector's magnetization being either parallel or antiparallel to the incoming spins (Fig. 3c). Upon reversing the bias current, $I_{dc}$ = -5 µA, the resulting hysteresis is mirrored with respect to B = 0 T (Fig. 3d). Such behavior has likewise been observed in experiments directed toward electrical detection of spin-polarized currents in 3D TIs[15-21]. This finding is consistent with either the injection of spin-polarized current from the BTS surface states into the graphene, or the spin Hall effect due to proximity-induced SO coupling[38], as both could generate a spin current that is perpendicular to the direction of the charge current imposed onto the BTS. In order to discriminate between the two mechanisms, we performed further experiments addressing the influence of three different parameters, namely the back gate, the temperature, and the thickness of the BTS nanoplatelet. The gained results together strongly favor the direct spin injection scenario, as detailed in the following.

Firstly, the spin signal was found to be only little affected and to lack a meaningful physical trend upon application of back gate voltages of up to ±45 V (Fig. 4a). If the spin Hall effect were present, the vertical electric field should alter the strength of the interfacial Rashba SO coupling, and thus influence the generated spin imbalance. At the same time, the appreciable n-type doping of the BTS sheets accounts for the absence of a sizeable gate effect for the spin injection scenario, as the Fermi level can only be slightly shifted even for highest gate voltages[39]. Within the relevant model of 1D spin transport[36] the spin signal $\Delta R_{NL}$ is given by

$$\Delta R_{NL} = 4R_G \, exp\left(-\frac{L}{\lambda_G}\right)\left[\left(\frac{P_J\frac{R_C}{R_G}}{1-P_J^2}+\frac{P_{Co}\frac{R_F}{R_G}}{1-P_{Co}^2}\right)\left(\frac{P_J\frac{R_{C'}}{R_G}}{1-P_J^2}+\frac{P_{Co}\frac{R_F}{R_G}}{1-P_{Co}^2}\right)\right]\left[\left(1+\frac{2\frac{R_C}{R_G}}{1-P_J^2}+\frac{2\frac{R_F}{R_G}}{1-P_{Co}^2}\right)\left(1+\frac{2\frac{R_{C'}}{R_G}}{1-P_J^2}+\right.\right.$$
$$\left.\left.\frac{2\frac{R_F}{R_G}}{1-P_{Co}^2}\right)exp\left(-\frac{2L}{\lambda_G}\right)\right]^{-1},$$
$$(1)$$

where $P_J$ is the polarization of the injected spin current by the BTS, $P_{Co}$ is the polarization of the cobalt contact, $R_c$ and $R_{c'}$ are the contact resistances, $R_G$ is the graphene channel resistance, $L$ is the distance between injector and detector, and $\lambda_G$ is the spin diffusion length. $R_F = \rho_{Co}.\lambda_{Co}/A_{Co}$, where $\rho_{Co} = 115$ nΩ.m is the cobalt resistivity at low temperature, $\lambda_{Co} = 6.10^{-8}$ m is the spin diffusion length of cobalt, and $A_{Co}$ is the cobalt contact area. Using this model, we estimate the spin polarization $P_J$ at the BTS-graphene junction to be 10%.

Secondly, the resistance change associated with the switching shows a temperature characteristic similar to that of magnetotransport-related phenomena in BTS. As apparent from Fig. 3b, the signal decreases significantly with increasing temperature, such that it becomes difficult to detect above 15 K. In the case of WAL[40] in BTS, the phase coherence length displays a $T^{-n}$ dependence, where n takes the value of 0.5 when the top and bottom surfaces are decoupled[41]. In previous magnetotransport studies of CVD-grown BTS, we have observed WAL features associated with 2D surface states up to about 20 K, which is close to the 15 K mentioned above[39]. Furthermore, in studies of BTS and $(Bi_{0.53}Sb_{0.47})_2Te_3$ devices, where the TI serves as both spin-injector and spin transport channel, spin transport has been observed up to 20 K in BTS[17] and up to 10 K in $(Bi_{0.53}Sb_{0.47})_2Te_3$[18]. These findings have been attributed to an increasing impact of the bulk TI states with increasing temperature. For the present spin valve devices, the extracted temperature dependent spin signal $\Delta R_{NL}$ exhibits a $T^{-n}$ dependence with 0.4<n<0.7. This range, obtained from all measured devices, is in close correspondence to the above mentioned WAL dependence, providing additional support that the spin generation occurs within the 2D surface states. Fig. 4b illustrates the similar dependence of the spin signal and of the extracted phase coherence length, $L_\varphi$, in BTS, extracted from the WAL magnetoconductance measurements, analogous to the data published by Tang et al[18] (see also Supporting Information Fig. S3 for comparison of the temperature dependence to all published data up to date).

Thirdly, the spin signal, acquired at $I_{dc}$ = 5 µA for several different devices with different BTS thicknesses, is seen to scale inversely with the thickness of the BTS platelets (Fig. 4c). This trend underscores that the main contribution to the measured signal stems from the BTS surface states, as contributions from the bulk of the nanoplatelets become more significant with increasing BTS thickness, analogous to conclusions drawn from spin-polarized current detection experiments on $Bi_2Se_3$[15].

We also investigated the dependence of the spin signal on the applied DC bias current for spin injection via the BTS. Upon applying bias currents of up to ±20 µA, a signal maximum is observed at $I_{dc}$ = ±5 µA, as shown in Fig. 4d. Previous studies on TI-based spin valves found both linear[15,20] and non-linear[17,21] dependence of the measured spin signal on the applied bias. For those devices, bias currents on the order of μA to mA have been applied in order to induce spin polarization. By comparison, the BTS platelets in our devices have much smaller dimensions (6-15 nm thickness, 1-3 μm lateral size), which leads to substantially higher current densities. On this basis, we attribute the spin signal decrease at larger bias to Joule heating effects in the BTS nanoplatelet.

Finally, with the aim of determining the spin relaxation times, we performed Hanle measurements by applying current between the non-magnetic BTS/Au electrodes and detecting $\Delta R_{NL}$ via the magnetic Co electrodes, while applying an out-of-plane magnetic field. The detector magnetization is first aligned in-plane by an external magnetic field, followed by tilting the device by 90° with respect to the magnetic field axis. This leads to precession of the spins with the Larmor frequency, $\omega_L = (g\mu_B B\perp)/\hbar$, while they propagate toward the detector, as schematically illustrated in Fig. 5a. For zero B-field and positive bias, the spins arrive parallel to the detector and the signal exhibits a maximum. With increasing B-field the signal becomes proportional to cos ($\theta$), $i.e.$, the projection of the spin direction onto the direction of the detector's magnetization and $\Delta R_{NL}$ decreases. Like for the spin valve signal in Fig. 3c,d, reversing the bias polarity causes a flip of the Hanle curve (Fig. 5b). In order to extract the spin life times, we fitted the Hanle data to Eqn. 2:

$$R_{NL} \propto \int_0^\infty \frac{1}{\sqrt{4\pi Dt}} exp\left[-\frac{L^2}{4Dt}\right] cos(\omega_L t) \, exp\left(-\frac{t}{\tau_s}\right) dt, \qquad (2)$$

where $\tau_s$ is the spin lifetime, $L$ is the contact separation, and $D$ is the spin diffusion constant[35,42]. The fits yield spin lifetimes in the range of 90 - 110 ps, spin diffusion lengths on the order of 400 nm for spins propagating in the graphene channel, and corresponding spin diffusion constants between $1.2 \cdot 10^{-4}$ and $3.4 \cdot 10^{-4}$ m$^2$s$^{-1}$. Previous studies of spin current injected into graphene have found spin lifetimes of up to 6.2 ns in bilayer graphene at 20 K in the presence of an MgO tunnel barrier[43] and diffusion lengths on the order of several μm[3,4,44]. Additionally, quite large spin lifetimes of up to 65 ns[45] have been observed for conduction electrons by electron spin resonance (ESR) measurements, attributable to the absence of metallic contacts or substrate effects[46]. Like for the spin valve resistance change, $\Delta R_{NL}$, a fast decay with increasing temperature also occurs for the amplitude of the Hanle curves, which are detectable only up to 10 K (Fig. 5d). Nonetheless, the diffusion length remains almost constant up to this temperature, which proves that the fading amplitude of the Hanle curves is solely due to diminished spin polarization within the BTS, rather than processes within the graphene itself.

Hanle curves measured for comparison using the Co injector (Fig. 5c) reflect similar diffusion lengths on the order of several hundreds of nm, along with a significant signal decrease with increasing temperature. In principle, the Hanle signal should persist up to higher temperatures[34]. However, two factors need to be taken into account in this respect. Firstly, the graphene quality is significantly decreased due to the multiple fabrication steps that leave behind surface contaminations on the graphene. Secondly, the Co contacts are strongly coupled to the graphene, which has previously been considered as an impediment in Hanle experiments. In general, for transparent contacts, appreciable contact-induced spin relaxation and consequently shorter lifetimes on the order of 100 ps[34,47] have been reported.

In summary, it follows that intimate coupling between a 3D TI and graphene allows for exploiting the SO-induced spin-momentum locking in the TI for injecting spin polarized currents into the adjacent graphene sheet. Our findings represent a major advancement toward the realization of novel spintronic device designs for energy efficient spin-logic applications. Moreover, the demonstrated devices open perspectives for realizing integrated circuitries combining graphene-based spin transport channels with logical elements. We anticipate that further improvement of the cleanliness of the graphene can enhance the performance of the hybrid spintronic devices. Importantly, due to the non-magnetic character of the TI injectors, the present devices open up novel, intriguing prospects for the development of all-electric spintronics. In future studies, the $Bi_2Te_2Se$ system could be replaced by a TI with improved properties, such as the tetradymite $Bi_{1.5}Sb_{0.5}Te_{1.7}Se_{1.3}$, whose Fermi level position favors dominating surface state contribution[48].

**Figures**

**a**

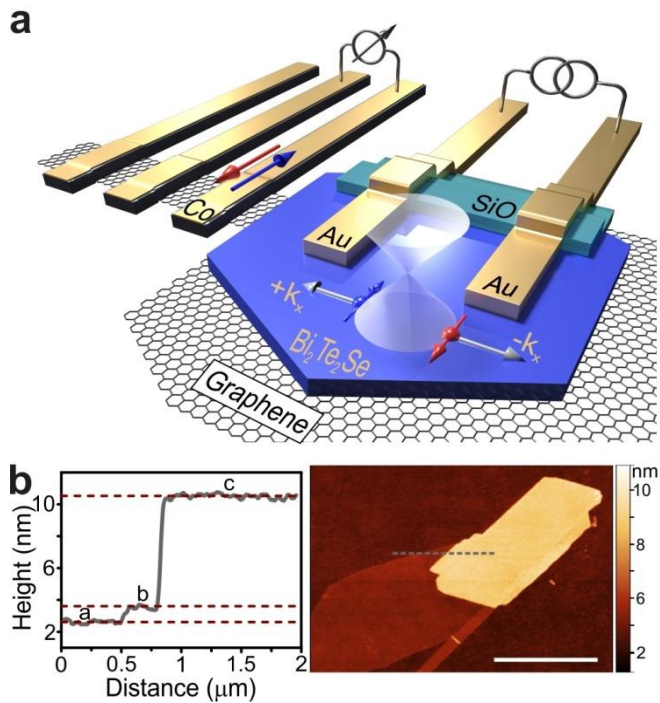

**Figure 1.** Experimental concept and device structure. (a) Schematic depiction of the investigated device configuration, with a graphene-TI heterostructure as major component. Due to its spin-momentum coupling, the TI is expected to act as spin polarizer on top of the graphene. The direction of the spin current is determined by the polarity of the applied bias and is detected by a magnetic electrode on top of the graphene channel. (b) AFM image of a graphene-$Bi_2Te_2Se$ nanoplatelet stack on $Si/SiO_2$ substrate **-a-**, obtained by van der Waals epitaxy growth of the TI **-c-** on top of mechanically exfoliated graphene **-b-**. The thickness of as-grown platelets ranges between 4 and 15 nm. (Scale bar: 2 µm)

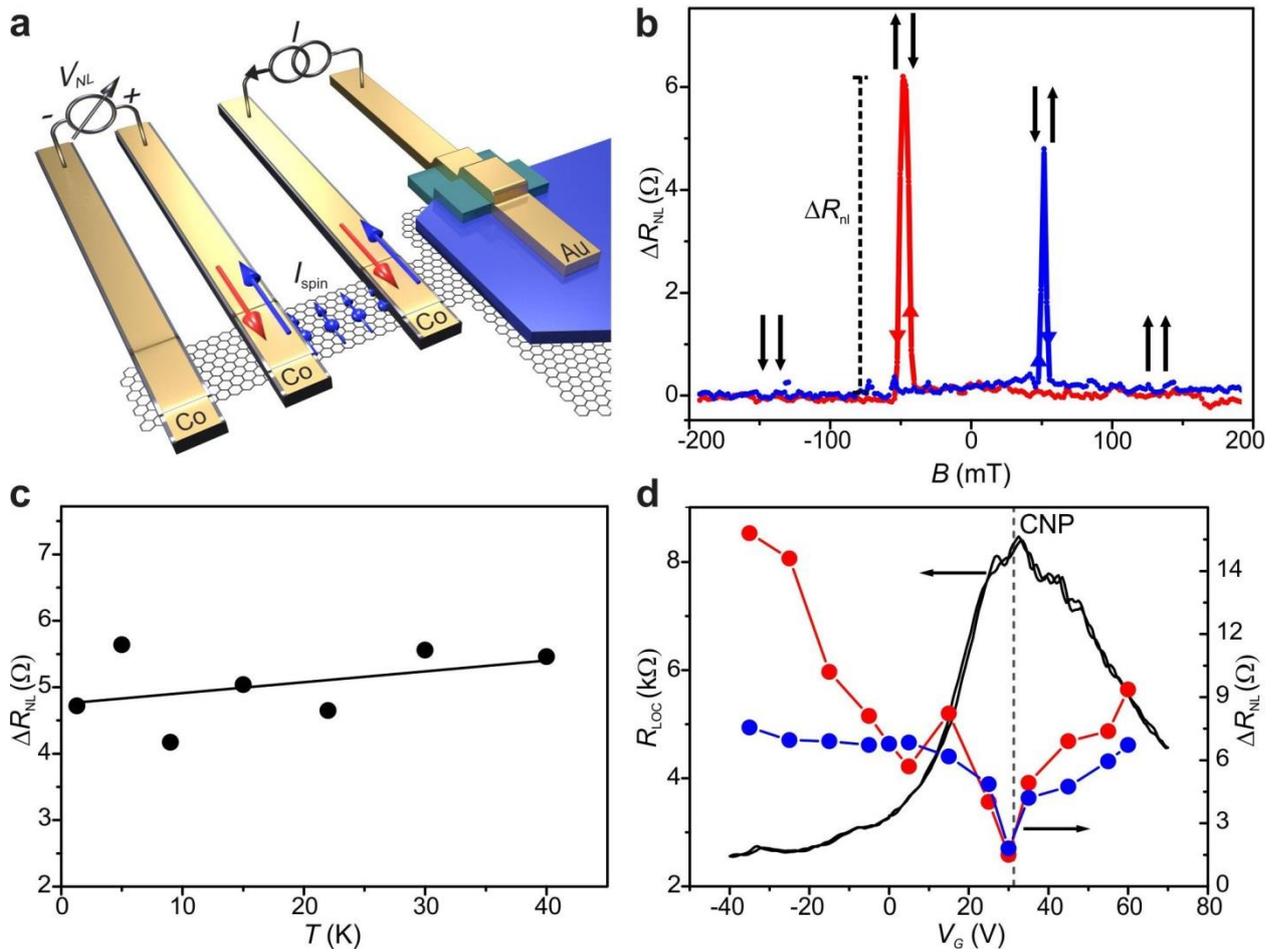

**Figure 2.** Spin injection into graphene by Co contacts. (a) Schematic diagram of the spintronic device with 4-terminal non-local measurement geometry. In the present experiments, a magnetic electrode injects spin-polarized carriers into the graphene channel (first Co contact is used as a source, the Au contact as drain), while the neighboring magnetic electrode detects it as a change in the non-local signal (second and third Co electrodes are the voltage probes). (b) Spin valve effect in graphene, where the non-local resistance change is plotted as a function of in-plane B-field strength. The resistance is low when the injector and detector electrode have the same direction of magnetization, and higher resistance when their magnetizations are antiparallel, as indicated by the arrows. (c) Non-local resistance change as a function of temperature. The spin valve signal is almost independent of temperature, in accordance with previous studies of graphene-based spin valves. This behavior is due to the fact that electron-phonon interactions in graphene are suppressed even at higher temperatures.

(d) A local resistance measurement reveals graphene's charge neutrality point (CNP) at $V_g =$ 32 V (black curve), consistent with p-type doping during the fabrication process. The gate dependence of the non-local spin valve signal has a minimum close to the CNP (blue and red curve for right and left jump respectively), indicating that the cobalt contacts are nearly transparent.

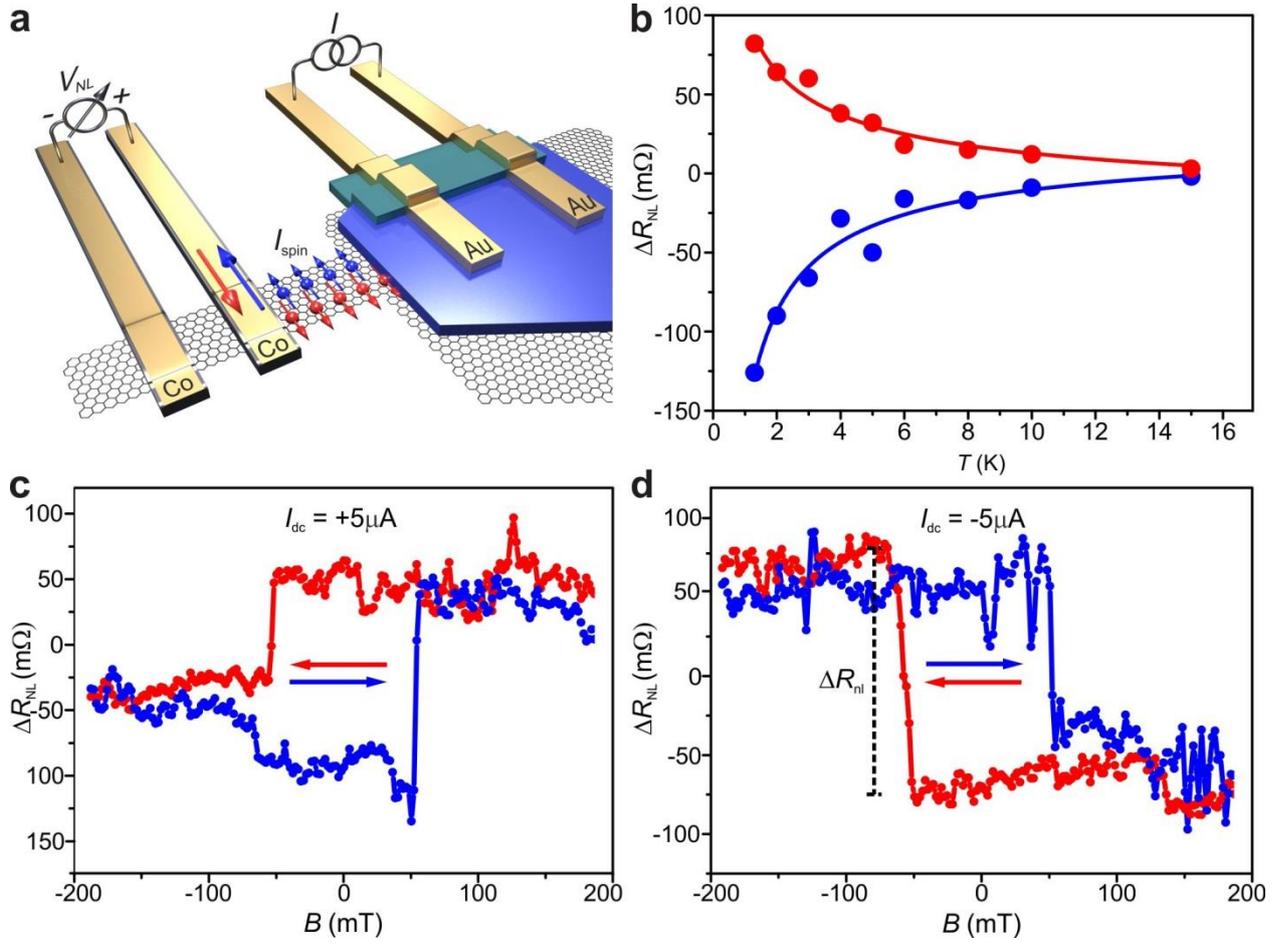

**Figure 3.** Electrical detection of spin currents in graphene injected from $Bi_2Te_2Se$ as spin polarizer. (a) Schematic diagram of the spintronic device with 4-terminal non-local geometry. In the present measurement configuration, the net spin current generated in the TI is injected into the graphene channel in a direction depending on the polarity of the applied bias (the left Au electrode serves as a source, the right one as drain). A magnetic Co electrode on the graphene strip serves as spin current detector. (b) Plot of the spin valve non-local signal in dependence of temperature for positive (blue) and negative (red) B field. The data closely follows a $T^{-n}$ dependence, where n = 0.7 and n = 0.5 respectively. (c) Spin valve non-local signal, detected as electrical resistance change upon reversing the magnetization direction of the magnetic detector electrode with respect to the incoming spin current. The measurement is taken at a bias current of $I_{dc}$ = +5 μA, for which maximum signal was observed. (d) For reversed sign of applied bias, the hysteresis is mirrored proving that the TI injects spin-

polarized current into the graphene channel. No linear magnetoresistance background has been subtracted from the data.

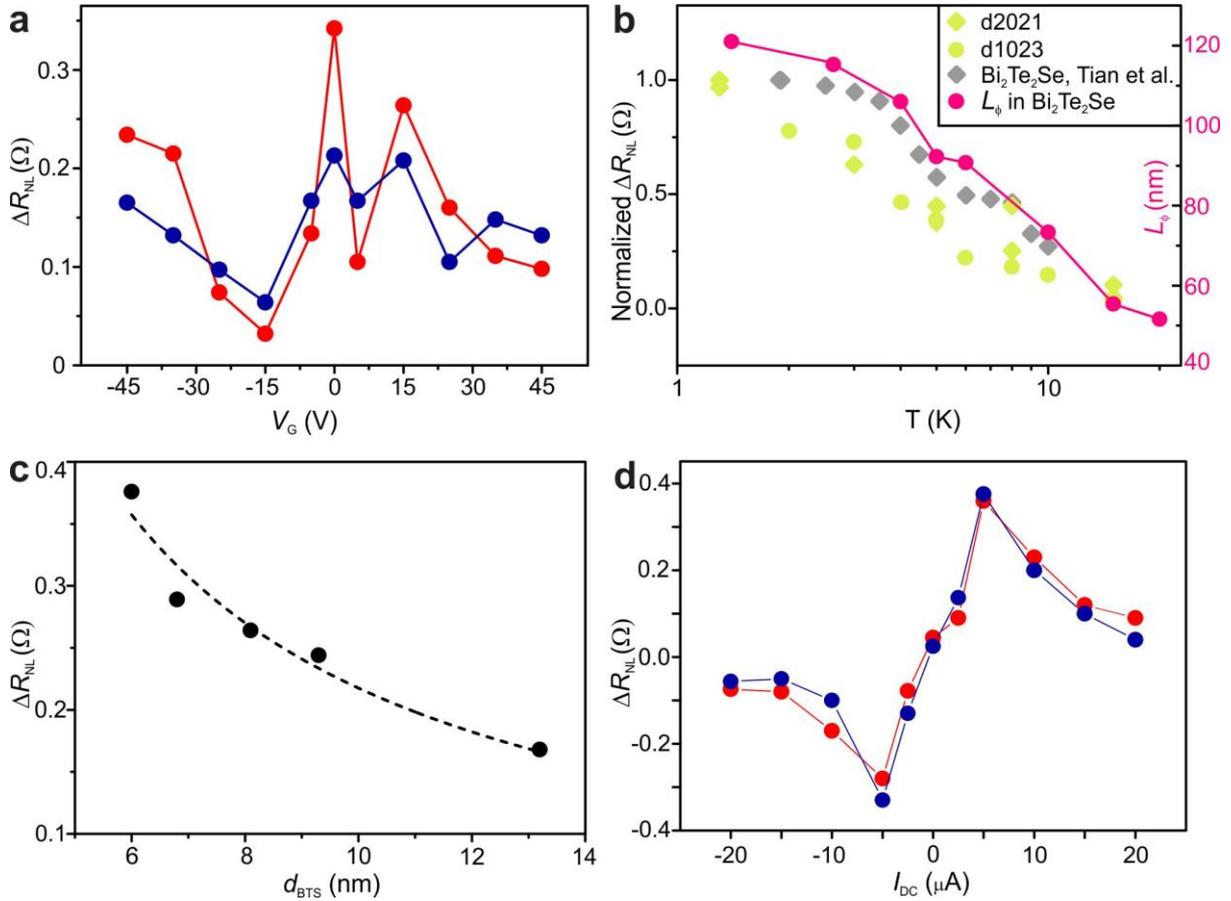

**Figure 4.** Spin signal dependencies. (a) No sizeable dependence of $\Delta R_{NL}$ on the back gate voltage in the range of ±45 V applied across the 300 nm SiO$_2$ dielectric layer for positive (blue) and negative (red) B-field. (b)Temperature dependences of $\Delta R_{NL}$ and the phase coherence length, extracted from magnetoconductance measurements of BTS, exhibit similar trend. (c) $\Delta R_{NL}$ acquired at $I_{dc}$ = +5 μA for five different BTS-platelet thicknesses, which scales inversely with the thickness, suggesting that the measured spin signal is very likely to originate from the BTS surface states. (d) $\Delta R_{NL}$, as a function of bias current, $I_{dc}$ in the range ± 20 μA. A maximum is observed for $I_{dc}$ = ± 5 μA, after which the signal decreases for increasing bias.

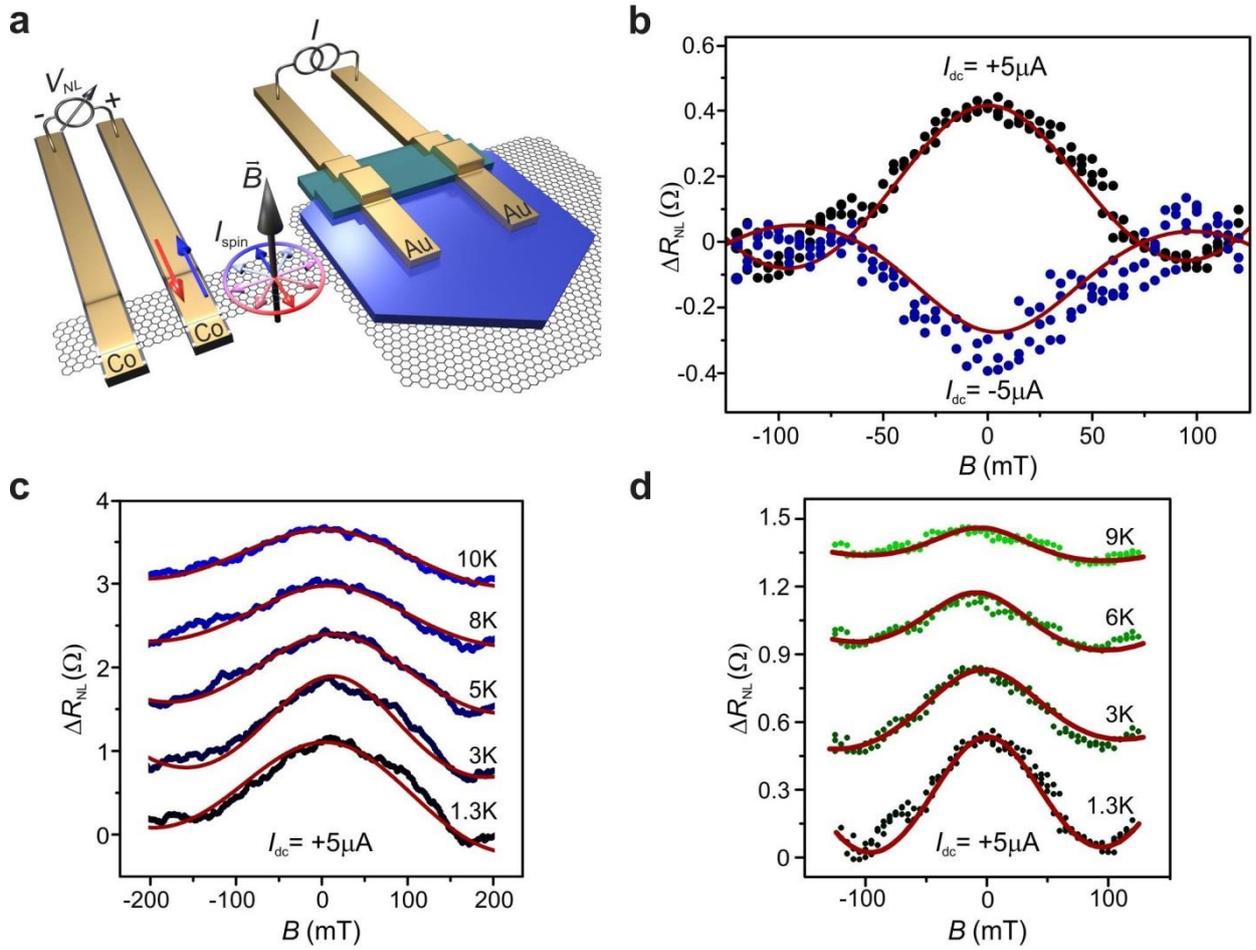

**Figure 5.** Hanle precession. (a) Hanle effect is induced in the graphene channel, when an external out-of-plane magnetic field is applied. The spins injected by the TI and transported in the graphene start precessing around this field and thus arrive at the detector electrode having a different orientation. The detected signal has a maximum at B = 0 T and decreases as the field increases and reorients the spins. (b) Low temperature (T = 1.3K) Hanle data for positive and negative bias where changing the bias current direction flips the Hanle peak. The solid lines are fits to the data using Eqn. 2 which yield a spin life time of 107 ps and a diffusion length of 430 nm. (c) Temperature dependence of the Hanle curves measured for the Co injector geometry. The similarly small diffusion constant and spin life time are attributed to the resist residues on the graphene after completing multiple fabrication steps. (d) Temperature dependence of the Hanle curves measured for the BTS/Au-injector

geometry, where the spin signal is seen to decrease as the sample temperature approaches 10 K. The solid lines are fits to the data and the curves are offset for clarity.

## ■ ASSOCIATED CONTENT

**Supporting Information**.

Contact resistances, AFM of a representative device, Raman spectroscopy of the BTS/graphene heterostructure; temperature dependence of the spin signal in comparison to literature. This material is available free of charge via the Internet at http://pubs.acs.org.

## ■ AUTHOR INFORMATION

**Corresponding Author**

*K.V.: E-mail: k.vaklinova@fkf.mpg.de

**Author Contributions**

K.V., A.H. and M.B. conceived and designed the experiments. K.V. grew the $Bi_2Te_2Se$-graphene heterostructures and fabricated the devices. K.V. and A.H. performed the electrical transport measurements and analyzed the data. K.V. and M.B. wrote the manuscript. K.K. supervised the overall project. All authors have contributed to the manuscript and have given approval to its final version.

**Funding Sources**

This work was supported by the Deutsche Forschungsgemeinschaft (DFG) within the framework of the priority program SPP1666.

**Notes**

The authors declare no competing financial interest.

## ■ ACKNOWLEDGMENTS

The authors acknowledge the Nanostructuring Lab Group at MPI, Stuttgart for providing the clean room facilities and M. Hagel for performing RIE.

## ■ REFERENCES

**Table of Contents Graphic**

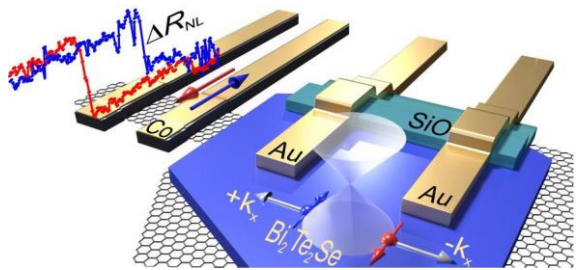